\documentclass[aps,english,prb,reprint,superscriptaddress]{revtex4-1}

\usepackage{graphicx}
 \usepackage{float}
\usepackage{amssymb}
\usepackage{soul}
\usepackage{babel}
\usepackage{color}
\usepackage{esvect}
\usepackage{amsmath}
\usepackage{amsfonts}

\begin{document}

\title{Ferromagnetic phase of spinel compound MgV$_2$O$_4$ and its spintronics properties}

\author{Javad G. Azadani}

\author{Wei Jiang}
\email{jiangw@umn.edu}

\author{Jian-Ping Wang}

\author{Tony Low}
\email{tlow@umn.edu}
\affiliation{Department of Electrical and Computer Engineering, University of Minnesota, Minneapolis, Minnesota 55455, USA}

\begin{abstract}
Spinel compound, MgV$_2$O$_4$, known as a highly frustrated magnet has been extensively studied both experimentally and theoretically for its exotic quantum magnetic states. However, due to its intrinsic insulating nature in its antiferromagnetic (AFM) ground state, its realistic applications in spintronics are quite limited. Here, based on first-principles calculations, we examine the ferromagnetic (FM) phase of MgV$_2$O$_4$, which was found to host three-dimensional flat band (FB) right near the Fermi level, consequently yielding a large anomalous Hall effect (AHE, $\sigma \approx 670\,\Omega^{-1}\cdot cm^{-1}$). Our calculations suggest that the half-metallicity feature of MgV$_2$O$_4$ is preserved even after interfacing with MgO due to the excellent lattice matching, which could be a promising spin filtering material for spintronics applications. Lastly, we explore experimental feasibility of stabilizing this FM phase through strain and doping engineering. Our study suggests that experimentally accessible amount of hole doping might induce a AFM-FM phase transition. 

\end{abstract}

\maketitle

\section*{Introduction}

Spinel compounds, with a large family of interesting magnetic materials, have been extensively studied for decades for their rich magnetic \cite{valenzuela2012novel,marco2001cation,phanichphant2012cellulose, zhao2017spinels}, electronic \cite{cho2011spinel,naveen2015cobalt}, optical \cite{sonoyama2006electrochemical}, and topological properties \cite{Xu2011,jiang2020magnetic}. In general, spinel compounds with chemical formula A$_{1-\alpha}$B$_{\alpha}$(A$_{\alpha}$B$_{2-\alpha}$)X$_{4}$ are classified into the normal ($\alpha = 0$), inverse ($\alpha = 1$), and complex (0$<\alpha<$1), depending on the different cations (A and B) distributions in octahedral and tetrahedral locations. The most widely investigated spinels are normal spinels with chemical formula AB$_{2}$X$_{4}$, where the A and B metal cations are located at the tetrahedral and octahedral sites, respectively. The conventional unit cell is presented in Fig. 1(a), where X ions form the face-centered cubic lattice and cations A and B form diamond and three-dimensional (3D) kagome sublattices, respectively. 

For the system to be charge neutral, anion X always accept two electrons while cations A normally donates two or four electrons and cations B donates two or three electrons, leading to the two possible {A}$^{4+}${B}$^{2+}_{2}${X}$^{2-}_{4}$ or {A}$^{2+}${B}$^{3+}_{2}${X}$^{2-}_{4}$ states. Therefore, with combinations of different cations A and B, we are able to selectively populate different sublattices. For example, within vanadium spinels, {V}$^{4+}${Mg}$^{2+}_{2}${O}$_{4}$ shows interesting magnetic Weyl semimetal state due to the diamond lattice constructed by V$^{4+}$ ions \cite{jiang2020magnetic}; while Mg$^{2+}$V$_2^{3+}$O$_4$ is a well known highly frustrated magnet that hosts intriguing quantum magnetic states due to the corner-sharing 3D kagome lattice \cite{wheeler2010spin}. Both of these spinels, in principle, could be experimentally synthesized by properly controlling the oxygen pressure and stoichiometric ratio between the elements \cite{oshima1980phase,hellmann1983,islam2012growth,niitaka2013type}. Nevertheless, the commonly observed phase experimentally is Mg$^{2+}$V$_2^{3+}$O$_4$, which is intrinsically an anti-ferromagnetic (AFM) insulator \cite{wheeler2010spin}. 
Studies on other spinel compounds, such as ZnCr$_2$O$_4$ and CaCr$_2$O$_4$, suggest hole doping can stabilize the ferromagnetic (FM) phase \cite{zhang2006local,dutton2011sensitivity,dutton2010divergent,onoda1997spin}. 
However, the electronic properties of the MgV$_2$O$_4$ in its FM phase is not well understood.
Here we reveal that the FM phase can accommodates interesting topological and spintronics properties. This includes large anomalous Hall conductivity, half metallicity, and AFM-FM and metal-insulator transitions. We also explore possible strain and doping approaches where this FM phase might be stabilize, which can motivate its experimental search.


In this work, we revisit the electronic and magnetic properties of MgV$_2$O$_4$ with first-principles calculations based on density functional theory (DFT). A significant point of departure of our work to prior literature is the emphasis on its 3D kagome lattice, and its characteristic flat band (FB). FBs are well studied in the literatures for their interesting physics such as superconductivity \cite{kobayashi2016superconductivity,miyahara2007bcs}, ferromagnetism \cite{mielke1991ferromagnetism,mielke1992exact,zhang2010proposed,hase2018possibility} and topological states \cite{tang2011high,Liu2013,Su2018,Jiang2019c,zhou2019weyl}. The FM MgV$_2$O$_4$ shows interesting FB feature and ideal half-metallicity with a large spin gap, which is further analysed based on tight-binding simulation of the 3D kagome lattice model. We found that intrinsic AFM MgV$_2$O$_4$ could transition into a FM state via experimentally accessible hole doping, such as ionic gel gating. Interestingly, the AFM to FM transition is accompanied with an insulator to metal transition and a giant variation of anomalous Hall effect. Lastly, we demonstrated how lattice and chemically matched interface between MgV$_2$O$_4$ and MgO could help to preserve its half-metallicity. These new predictions could potentially usher in future novel and practical spintronics applications.

\section*{First-principles calculation results}
\begin{figure}
\includegraphics[width=8.5cm]{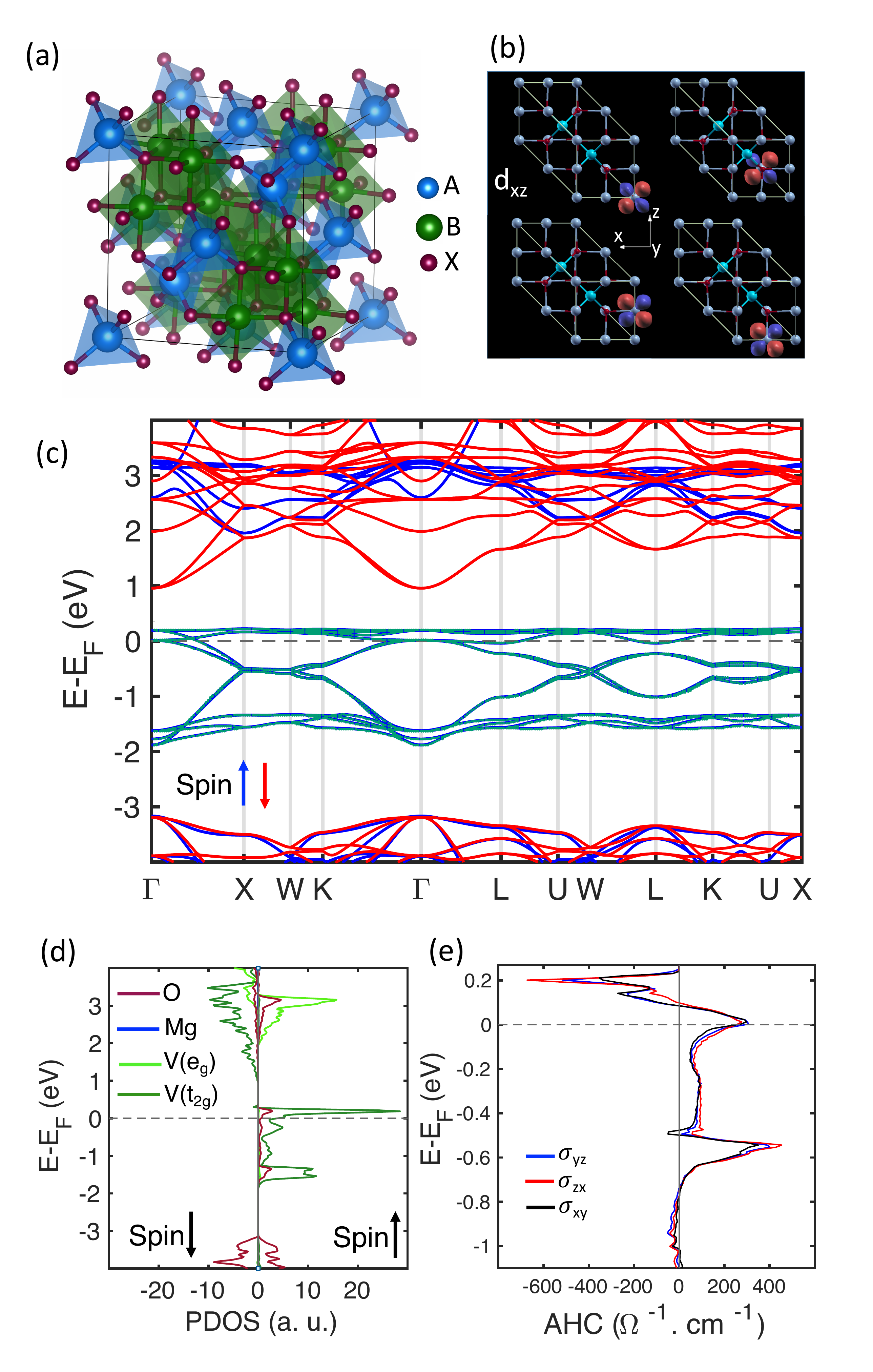}
\caption{ First-principles calculations. (a) Conventional crystal structure of spinel compound with formula AB$_{2}$X$_{4}$, where metals A and B occupy the centers of tetrahedrons (blue) and octahedrons (green), respectively. (b) Maximally localized Wannier functions showing d$_{xz}$ orbitals of four vanadium atoms in the lattice. (c) Band structure of MgV$_{2}$O$_{4}$ in its FM state, where blue and red colors corresponds to the spin-up and spin-down channels, respectively. MLWF fitted bands is overlaid with green dotted lines. (d) Projected density of states (PDOS) of MgV$_{2}$O$_{4}$, showing the strong contribution of t$_{2g}$ orbitals of vanadium atoms around the Fermi level. The zero of the energy is set to the Fermi energy. (e) Anomalous Hall conductivity (AHC) plotted, showing large peaks of 670\,$\Omega^{-1}\cdot cm^{-1}$ and 450\,$\Omega^{-1}\cdot cm^{-1}$ at 0.20\,eV above and 0.54\,eV below the Fermi energy, respectively. The magnetization direction is along (111). }
\label{FIG1}
\end{figure}

We start by studying the electronic properties of MgV$_{2}$O$_{4}$ in the FM state rather than commonly studied AFM phase (Supplemental Material \cite{Supp}). To treat the strongly localized $d$ electrons of V ions, an on-site Hubbard U term is added. Using the documented value (U=3.0\,eV) \cite{jain2013commentary,pandey2011orbital}, spin-polarized band structure for FM MgV$_{2}$O$_{4}$ is presented in Fig.~\ref{FIG1}(c), which interestingly shows an ideal half-metallic feature with a large band gap of 4.15\,eV for the spin-down channel. The half-metallicity will remain for different U values, which only changes the separation between two spin channels \cite{Supp}. For example, in Fig. 1(c), the minority spin (red) bands are separated from the majority spin (blue) Fermi surface by about 1\,eV. Moreover, MgV$_{2}$O$_{4}$ possesses four nearly FBs right above the Fermi energy, with their bandwidths varying from 29\,meV (very flat) to 205\,meV. Detailed analysis show that there are two three-fold degenerated points formed by two FBs and one dispersive band at the center of Brilluoin zone (BZ), ($\Gamma$) at energies 16\,meV and 196\,meV above the Fermi level, respectively.

To reveal the nature of these FBs, we calculated orbital and orbital-projected density of states (PDOS), as shown in Fig.~\ref{FIG1}(d). Around the Fermi level, there is a sharp peak that corresponds to the FBs. These FBs are mainly contributed by spin up electrons of the vanadium atoms, which form the 3D kagome lattice. This agrees with the magnetic moment distribution, which are mainly localized on V atoms. Detailed analysis of PDOS shows that all the twelve bands near the Fermi level (E$_{\mathrm{F}}$- 1.88 to E$_{\mathrm{F}}$+ 0.22\,eV) are contributed by the $t_{2g}$ orbitals. This can be understood from the crystal field splitting of the V atoms in the octahedron formed by neighboring O ions, which splits the five degenerate $d$ orbitals into two $e_g$ and three $t_{2g}$ orbitals with relatively lower energy. Due to the slight distortion of the octahedron, three degenerate $t_{2g}$ orbitals are further splitted into one lower energy $d_{xy}$ and two nearly degenerate $d_{xz}$ and $d_{yz}$ orbitals, which is further validated by the different O-V bond angle in the octahedron \cite{Supp}. After donating two $s$ and one $d$ electrons, V$^{3+}$ ions have two $d$ electrons left. Therefore, with the remaining two valence electrons of V$^{3+}$, the $d_{xy}$ orbitals with the lowest energy will be fully filled while the other two degenerate $d_{xz}$ and $d_{yz}$ orbitals will be half-filled.

In forming the band structure, each $d$ orbitals will, in principle, form one set of 3D kagome lattice, and the band filling will be determined by the number of valence electrons of V. These indeed agree perfectly with the band structure from DFT, where the bottom set of 3D kagome bands are fully occupied and the upper eight bands are half filled, which correspond to the two sets of 3D kagome bands. We note that the large difference of the band width between those 3D kagome bands is due to the distortion induced changes of the hopping $t$ for different $d$ orbitals. We also performed maximally localized Wannier functions (MLWFs) calculations using the \textsc{Wannier90} package \cite{marzari1997maximally}. As shown in Fig. 1(c), the MLWF fitted band structure agrees perfectly with the DFT band structure. The plotted MLWFs further validate the orbital characteristics of vanadium atoms [see Fig.~\ref{FIG1}(b)].

With the increasing interest in spintronics, materials with ideal half-metallicity and large spin/anomalous Hall effect are extensively studied. However, materials with both of those features are still quite rare. The ideal half-metallicity of the FM MgV$_{2}$O$_{4}$ has already been demonstrated from the electronic band structure, which shows a spin gap as large as 4.15\,eV. Further, we study the anomalous Hall effect by calculating the anomalous Hall conductivity (AHC). First, we recalculate the band structure of MgV$_{2}$O$_{4}$ by turning on the spin-orbit coupling (SOC) and then fit the results using MLWFs, based on which a tight-binding Hamiltonian is obtained \cite{Supp}. The AHC can be acquired by integrating the Berry curvature of the occupied bands. We plotted the energy-dependent AHC of MgV$_{2}$O$_{4}$, as shown in Fig. 1(e). It can be clearly seen a large AHC peak of 670\,$\Omega^{-1}\cdot cm^{-1}$ at 0.20\,eV above the Fermi energy is observed, which can be traced to the 3D FBs. There is also a large AHC of 450\,$\Omega^{-1}\cdot cm^{-1}$ at energy 0.54\,eV below the Fermi energy. Note that the three components of AHC have relatively similar values, which is related to the direction of magnetization, that is along (111) direction. The large AHC can arise from the large Berry curvatures and electronic topology of the 3D kagome lattice, which only present in the FM state of MgV$_{2}$O$_{4}$. Detailed analysis about the AHC is beyond the scope of this work, which will be discussed in a separate study \cite{Jiang2020GAHE}.

\section*{Tight-binding analysis}
\begin{figure}
\includegraphics[width=8.9cm]{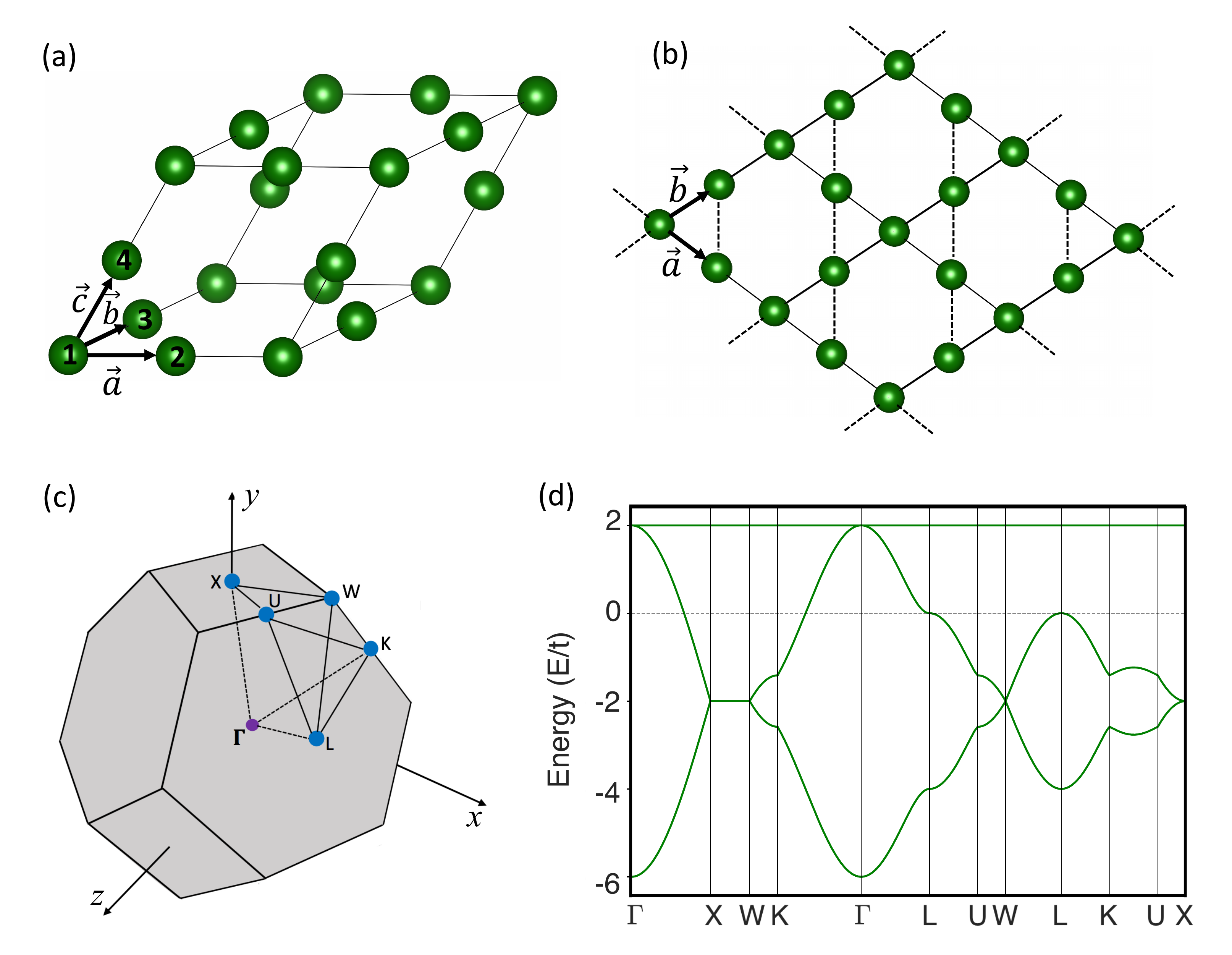}
\caption{Tight-binding analysis on 3D Kagome lattice. (a) Primitive unit cell of the pyrochlore lattice, where only atoms which form a 3D Kagome lattice are shown. (b) Cross section along each of the four faces of the tetrahedron produces one 2D kagome lattice, which one of them (\textit{a-b} plane) is shown here. (c) The first Brilluoin zone of the 3D kagome lattice with high-symmetry points and lines indicated. (d) Tight-binding band structure of 3D Kagome lattice, containing two-fold degenerate flat bands at $E$ = 2$t$, where $t$ is hopping amplitude for NN interactions.}
\label{FIG2}
\end{figure}

Since the bands of interest are mainly contributed by vanadium $d$ electrons, we can better understand the formation of the FB by analyzing the tight-binding model on a 3D kagome lattice. The primitive unit cell of the 3D kagome lattice contains four atomic states, with one corner-site and three edge-center-site states as presented by green atoms (B) in Fig.~\ref{FIG2}(a). Here, $\vec{a}$, $\vec{b}$, and $\vec{c}$ are the three primitive vectors pointing between nearest-neighbor (NN) lattice points that form a tetrahedron. It can be seen that the cross section along each of the four faces of the tetrahedron produces one 2D kagome lattice, as shown in Fig.~\ref{FIG2}(b), and for this we dubbed the system 3D kagome lattice. The first BZ of this lattice with high-symmetry lines is shown in Fig.~\ref{FIG2}(c). For simplicity, we consider one state per lattice site and only take NN interactions into account in our model. The spinless Hamiltonian is written as
\begin{equation}
H = \sum_{}^{} \epsilon_{i}c_{i}^{\dagger}c_{i} +  \sum_{<ij>}^{} t c_{i}^{\dagger}c_{j} + H.c.,
\end{equation} 
where $\epsilon_{i}$ is the on-site energy that is set as zero, and $\langle ij \rangle$ denotes summation over NN sites. $c_{i}^{\dagger}$ ($c_{j}$) creates (annihilates) an electron on the site $\vv{r_i}$ ($\vv{r_j}$), and NN hopping $t$ are set as the same value for all the NNs considering the structural symmetry.

By diagonalizing the Hamiltonian, we obtain the analytical energy bands as $E_{1,2} = -2$t$(1\pm \sqrt{1+A})$ and $E_{3,4}$ = 2$t$. $A$ is given by
$A = \cos{(2k_{x})}\cos{(2k_{y})}+\cos{(2k_{x})}\cos{(2k_{z})}+\cos{(2k_{y})}\cos{(2k_{z})}$, where $\vec{k}$ = ($k_{x}$,$k_{y}$,$k_{z}$) is the reciprocal lattice vector. The band structure along high-symmetry k-paths is shown in Fig.~\ref{FIG2}(d), which shows clearly the completely dispersionless FBs with double degeneracy located on top of two dispersive bands. The FBs touch the dispersive band at the $\Gamma$ point, forming a triple degenerate point, and the two dispersive bands construct the nodal lines along diagonal directions of the square face of BZ (i.e. X-W line). These agree with the calculated band structure of MgV$_{2}$O$_{4}$. When considering two $d$ orbitals with slightly different on-site energy, we are able to reproduce the DFT band structure \cite{Supp}.

\section*{Metal to insulator transition}

However, MgV$_2$O$_4$ is known to have a frustrated AFM ground state, which possesses an insulating state with different electronic properties than the FM case \cite{anderson1956ordering,mamiya1997structural,pandey2011orbital}. In contrast to the FM state, 3D FBs are absent in the AFM ground state of MgV$_{2}$O$_{4}$. We seek to engineer the magnetic ground state of MgV$_{2}$O$_{4}$ based on DFT calculations. It is worth mentioning that SOC and on-site Coulomb interaction are known to be important for the properties of MgV$_{2}$O$_{4}$ \cite{pandey2011orbital} and thus all are included in our calculations (see calculation details in Ref. \cite{Supp}). First, the total energy of the intrinsic MgV$_{2}$O$_{4}$ in the FM and AFM states are calculated, which shows that AFM energy, E$_{\mathrm{AFM}}$ is about 0.93\,eV lower than FM state, E$_{\mathrm{FM}}$, in agreement with previous studies \cite{niitaka2013type,wheeler2010spin}. There are several approaches to tuning the magnetic state, e.g., chemical doping, structural engineering, or external magnetic field~\cite{Enkovaara2003,White2013}. The magnetic field required to change the magnetic state depends on the exchange coupling strength, which is usually hard to manipulate and impractical for devices. Therefore, we focus on the chemical doping and lattice engineering methods.

\begin{figure}
\includegraphics[width=\columnwidth]{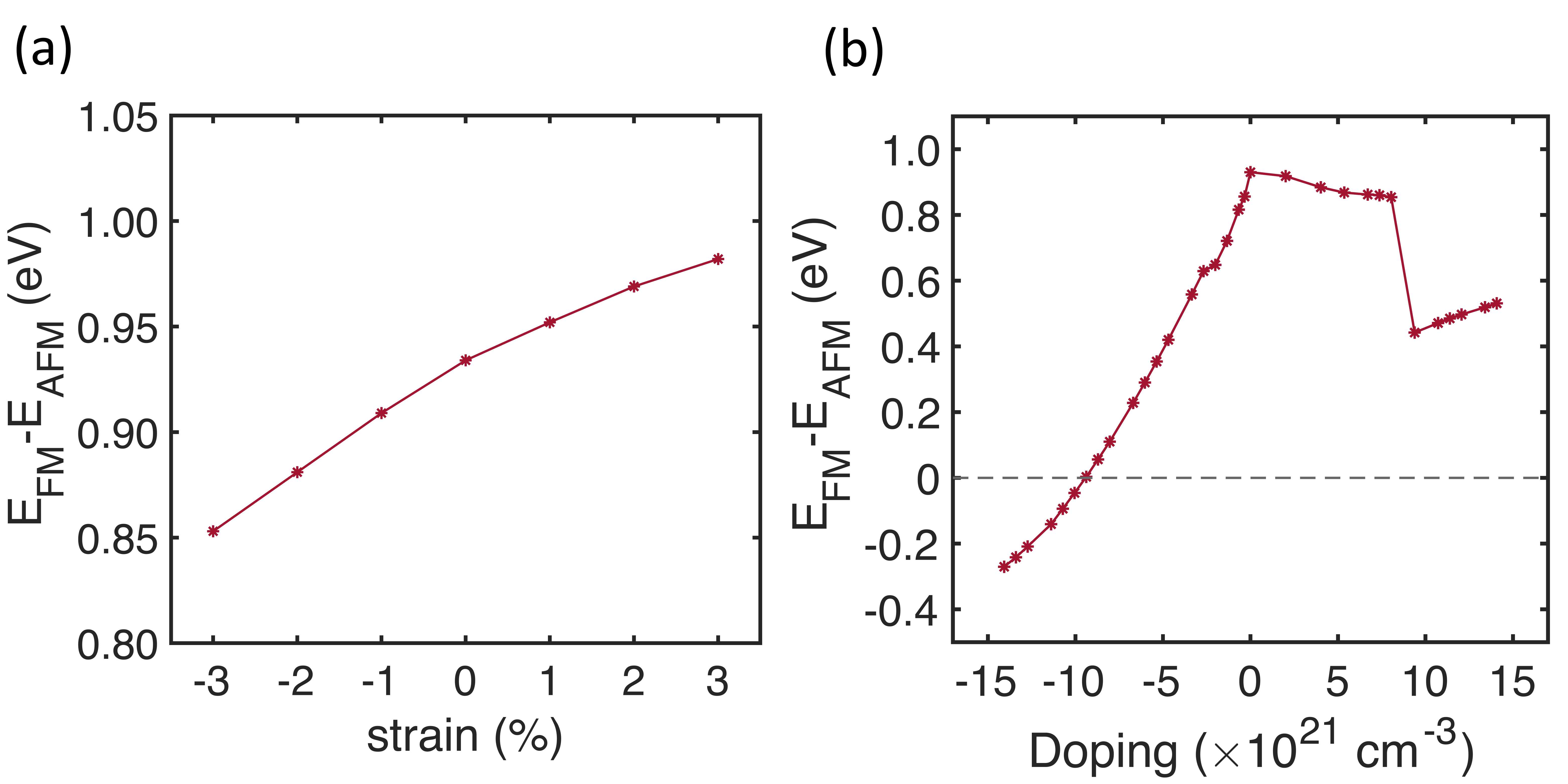}
\caption{ Engineering the magnetic ground state of MgV$_{2}$O$_{4}$. (a) Energy difference between FM and AFM states versus different values of anisotropic strain along y-direction. MgV$_{2}$O$_{4}$ remains to be AFM for all values of strains. (b) Energy difference between non-collinear calculations of FM and AFM states versus doping concentration, where positive and negative values corresponds to electron and hole doped MgV$_{2}$O$_{4}$, respectively. Hole doping of $10\times 10^{21}$ $cm^{-3}$ switches the magnetization of MgV$_{2}$O$_{4}$ to the FM.}
\label{FIG3}
\end{figure}

The strain engineering is accomplished through changing the lattice constant of MgV$_{2}$O$_{4}$. We calculated the E$_{\mathrm{FM}}$ and E$_{\mathrm{AFM}}$ after applying both isotropic and anisotropic strains. However, within the experimental accessible range, the ground state of MgV$_{2}$O$_{4}$ remains to be AFM, as shown in Fig. 3(a). To study the doping effect, we doped the system by changing the total number of electrons of the system while maintaining the charge neutrality with a compensating homogeneous background charge. E$_{\mathrm{FM}}$ and E$_{\mathrm{AFM}}$ for both hole and electron doping scenarios are calculated. As shown in Fig. 3(b), the system remains to be AFM for electron doping while changes to FM with the hole doping of $1\times 10^{22}$ $cm^{-3}$. We note that this doping concentration is experimentally feasible, which can be achieved through either element substitution or ionic gel gating methods \cite{yang2012doping,suh2014doping}. It is also important to mention that apart from a rigid shift of its Fermi level, characteristic band features are preserved after doping, i.e., the 3D kagome bands \cite{Supp}. Therefore, hole doping leads to an insulator-to-half metal transition for MgV$_{2}$O$_{4}$, which is consistent with previous experimental observations with Li doping of Li$_x$Mg$_{1-x}$V$_2$O$_4$ \cite{onoda1997spin}.

Considering the large variety of the spinel compounds, it should also be possible to find another compounds with FM ground state that host 3D FBs. Such hole doping induced AFM to FM transition has also been reported in chromium (III) spinels \cite{dutton2011sensitivity,dutton2010divergent}, which could possibly arise due to the breakdown of magnetic frustration, leading to the transition from long-range AFM coupling to short-range FM ordering. Considering the insulating feature of the AFM state without FBs, it is reasonable to expect a negligible AHC in the intrinsic MgV$_{2}$O$_{4}$ without hole doping. Therefore, through the hole doping, we actually accomplish a metal to insulator transition with a dramatic change of the AHE simultaneously. 
We note that due to the limitation of standard DFT in dealing with strong-correlated electrons, exact doping concentration to induce such phase transition from AFM to FM might varies, but the qualitatively phase transition should be valid as already demonstrated by related experiments \cite{mamiya1997structural}. Nevertheless, within the standard DFT formalism where electrons correlations are treated in a Hartree-Fock mean field manner, we conclude hole doping to be the most effective and feasible approach to experimentally access the intriguing FM state of MgV$_{2}$O$_{4}$.

\section*{Half-metallicity interface}
Finally, it should be noted that MgV$_{2}$O$_{4}$ has a very small lattice mismatch of $\sim$ 0.7\% with MgO, and a similar structure, yielding a smooth interface with MgO. Therefore, it could be used as a spacer layer in magnetic tunnel junction (MTJ) devices, as shown in Fig. 4(a), or as a spin filtering layer \cite{jiang2020magnetic}. To demonstrate this possibility, we directly simulate the MgO/MgV$_{2}$O$_{4}$/MgO heterostructure along (111) direction with three V layers.
Interestingly, MgV$_{2}$O$_{4}$ remains a half-metal even after integrating with MgO. Figure 4(b) presents total DOS of the heterostructure and PDOS of MgV$_{2}$O$_{4}$ layer in the supercell, which shows an almost complete spin polarization for the sandwiched MgV$_{2}$O$_{4}$ layer. 

\begin{figure}
 \includegraphics[width=9cm]{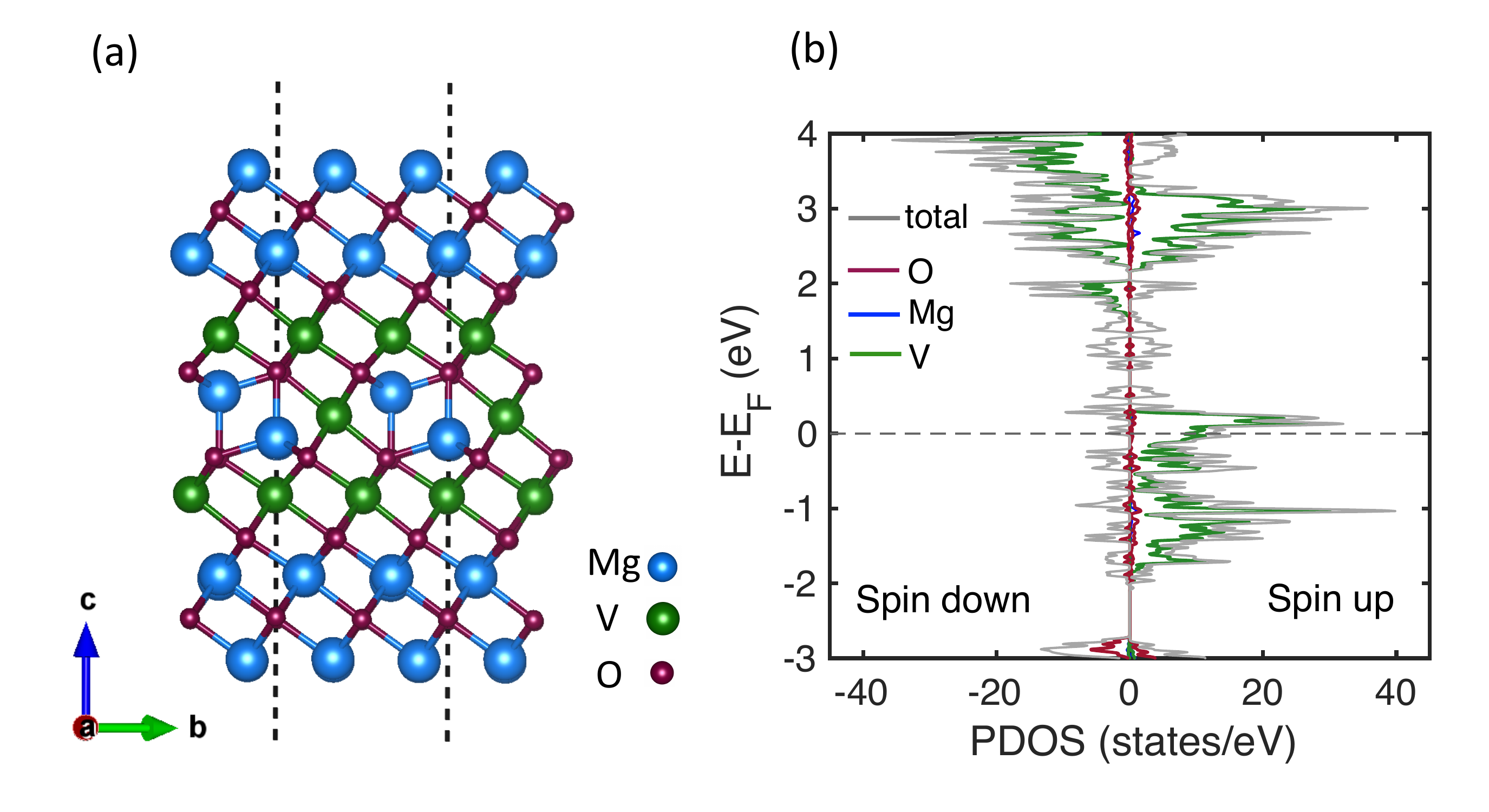}
\caption{ (a) MgO/MgV$_{2}$O$_{4}$/MgO supercell. (b) Projected DOS of spacer layer MgV$_{2}$O$_{4}$ and total DOS of MgV$_{2}$O$_{4}$ and MgO heterostructure, showing robust and complete spin polarization of MgV$_{2}$O$_{4}$ after integrating with MgO.}
\label{FIG4}
\end{figure}

\section*{Summary}
In this paper, by using first-principles calculations, we studied the experimentally synthesized MgV$_2$O$_4$ spinel compound and observed interesting 3D FBs in its FM phase. Vanadium atoms reside at the center of octahedrons forms a geometric frustrated 3D kagome lattice, which give rise to the 3D FBs right above the Fermi energy. PDOS calculations revealed that vanadium $t_{2g}$ orbitals contribute to the FBs, which was further confirmed by MLWFs calculations. Through hole doping, MgV$_{2}$O$_{4}$ with the insulating AFM ground state can be tuned into FM state, which is an ideal half-metal. Interestingly, the FBs contribute to a giant anomalous Hall effect right above the Fermi level. We believe these findings can greatly enrich the materials that have 3D FB, which could facilitate future experimental validation. Also, the FM-AFM (half-metal-insulator) transition with tunable AHE will also broaden the application of spinel compounds in spintronics.
 
\section*{Acknowledgements}
This work is supported by SMART, one of seven centers of nCORE, a Semiconductor Research Corporation program, sponsored by National Institute of Standards and Technology (NIST). We acknowledge computational support from the Minnesota Supercomputing Institute (MSI).





\bibliographystyle{unsrt}

\end{document}